\documentclass[a4paper,11pt]{amsart}
\begin{document}

\title{{\bf  On Michell-Laplace dark body}}
\author{Angelo Loinger}
\date{}
\address{Dipartimento di Fisica, Universit\`a di Milano, Via
Celoria, 16 - 20133 Milano (Italy)}
\email{angelo.loinger@mi.infn.it}

\begin{abstract}
The dark body of Michell-Laplace has nothing to do with the
relativistic black hole.
\end{abstract}

\maketitle

\vskip1.20cm
Many physicists believe that the notion of black hole (BH) is the
natural relativistic generalization of the Newtonian notion of
dark body (DB), investigated by Michell (1784) and by Laplace
(1796).
\par
Now, as it has been shown by McVittie \cite{1} (see also
Eisenstaedt \cite{2}), the above belief is based only on ``a play
of words in expressions such as \emph{the velocity of escape} or
\emph{the escape from a body}''. Since a recent article by
Stephani \cite{3} has given a further support to this conceptual
confusion, it is useful to recall here the very simple argument by
McVittie.
\par
Let us indeed consider a celestial spherical body of radius $R$
and mass $M$. According to Newtonian dynamics, the velocity of
escape $w$ of a particle, which is projected radially outwards
from the surface of the body, is given by

\begin{equation} \label{eq:one}
    w^{2}=2GM/R,
\end{equation}

where $G$ is the gravitational constant.\par If the particle is
projected with a velocity $u$ smaller than $w$, i.e. if

\begin{equation} \label{eq:two}
    u^{2}<2GM/R,
\end{equation}

it will arrive at a finite distance from the celestial body, and
then it will fall on its surface again. \par By using the
Newtonian corpuscular theory of light, which says that light is
composed of corpuscles obeying Newton's law of gravitation, and
travelling with a given velocity $c$, Michell and Laplace remarked
that if $c<w$ the light corpuscles cannot go away indefinitely
from the celestial body. Only if the radius $R$ of the body is
such that

\begin{equation} \label{eq:three}
    R=2GM/c^{2},
\end{equation}

they can escape from the gravitational attraction exerted by the
mass $M$. Then, if

\begin{equation} \label{eq:four}
    R<2GM/c^{2},
\end{equation}

the light corpuscles will attain a finite distance $d$ from the
celestial body, and an observer situated at an intermediate
distance between $R$ and $d$ will see the celestial body, owing to
the light corpuscles which arrive at his eyes.
\par If only the spherosymmetrical BH of GR existed -- in reality
this notion, and the notion of Kerr's BH, are the product of a
misinterpretation of the formalism of GR \cite{4}--, it ought to
have the fundamental property that neither the material particles
nor the light corpuscles can leave its surface. Therefore, such an
object would be invisible to any observer, however near he might
be. None of the phenomena observed by the experimentalists in the
region surrounding a DB of Michell-Laplace is present in the
neighborhood of a BH of GR.\par The imagined connection with the
Newtonian formula (\ref{eq:three}) comes out in this way: if for
the determination of the Einsteinian gravitational field generated
by a point mass $M$ (Schwarzschild's problem) one chooses the
\emph{\textbf{standard}} form -- as Droste, Hilbert and Weyl
(\emph{\textbf{but not Schwarzschild}} \cite{5}) did -- the radial
coordinate $r_{0}$ of the points of the space surface $r=r_{0}$
($r_{0}$ is the ``radius of the BH'') is given by

\begin{equation} \label{eq:five}
    r_{0}\equiv2GM/c^{2};
\end{equation}

this formula resembles the Newtonian formula (\ref{eq:three}),
which concerns an escape velocity $c$. But eq.(\ref{eq:three})
implies obviously that all observers -- including those at an
infinitely great distance from the celestial body -- can see it.
As it is clear, the Newtonian dark body of Michell-Laplace is
\emph{\textbf{not}} a black hole!

\small \vskip0.5cm\par\hfill {\emph{SCH\"ULER:}
  \par\hfill \emph{Kann Euch nicht eben ganz verstehen.}
  \par\hfill \emph{MEPHISTOFELES:}
  \par\hfill \emph{Das wird n\"achstes schon besser gehen,}
  \par\hfill \emph{Wenn Ihr lernt alles reduzieren}
   \par\hfill \emph{Und geh\"orig klassifizieren.}
   \vskip0.10cm\par\hfill Goethe, \emph{Faust (vv.1943-46)}

\end{document}